%% file: main.tex
\documentclass[letterpaper, 10 pt, conference]{ieeeconf}
\usepackage{times}
\usepackage{color, colortbl, xcolor}
\usepackage{todonotes}

\usepackage{multicol}
\usepackage[bookmarks=true]{hyperref}
\usepackage{amsfonts}
\usepackage{amssymb}
\usepackage{algorithmic}
\usepackage[ruled,vlined,linesnumbered,titlenotnumbered]{algorithm2e} 
\usepackage{amsmath}
\usepackage{dsfont}
\usepackage{multicol}
\usepackage{mathtools}
\usepackage{subcaption}
\usepackage{etoolbox}
\usepackage{balance}

\providetoggle{revision}
\togglefalse{revision}

\usepackage[%
    style=numeric-comp,
    sorting=none,
    backend=biber,
    sortcites=true,
    doi=false,
    firstinits=true,
    hyperref,
    isbn=false,
    eprint=false,
    minbibnames=10, 
    maxbibnames=10, 
    block=none]
    {biblatex}
    
\renewbibmacro{in:}{}
\AtEveryBibitem{%
  \clearlist{language}%
  \clearfield{pages}%
}

\bibliography{references}

\input{notation}

\graphicspath{{./figures/}}    
\DeclareGraphicsExtensions{.pdf,.png}

\IEEEoverridecommandlockouts
\usepackage[textfont=md,font=footnotesize]{caption}

\usepackage{ifthen}
\newboolean{include-notes}
\setboolean{include-notes}{true}

\usepackage{lipsum}

\newcommand{\dfknote}[1]%
    {\textcolor{orange}{\textbf{[DFK: #1]}}}
\newcommand{\ernote}[1]%
    {\textcolor{blue}{\textbf{[ER: #1]}}}
\newcommand{\jsnote}[1]%
    {\textcolor{purple}{\textbf{[JS: #1]}}}
\newcommand{\JS}[1]%
    {\textcolor{purple}{\textbf{[JS: #1]}}}
\newcommand{\adnote}[1]%
    {\textcolor{magenta}{\textbf{[AD: #1]}}}
\newcommand{\revision}[1]{\iftoggle{revision}{\textcolor{magenta}{#1}}{#1}}

\newcommand{\remove}[1]%
    {\textcolor{red}{#1}}

\newcommand{\example}[1]%
{
\textbf{Running example:}
\textit{#1}
}

\title{\LARGE \bf Efficient Iterative Linear-Quadratic Approximations\\for Nonlinear Multi-Player General-Sum Differential Games}

\author{
David Fridovich-Keil, Ellis Ratner, Lasse Peters, Anca D. Dragan, and Claire J. Tomlin
\thanks{
Department of EECS, UC Berkeley, \href{mailto:dfk@berkeley.edu}{\tt \small{\{dfk, eratner, lasse.peters, anca, tomlin\}@eecs.berkeley.edu}}.}%
\thanks{This research is supported by an NSF CAREER award, the Air Force Office of Scientific Research (AFOSR), NSF's CPS FORCES and VeHICaL projects, the UC-Philippine-California Advanced Research Institute, the ONR MURI Embedded Humans, a DARPA Assured Autonomy grant, and the SRC CONIX Center. D. Fridovich-Keil is supported by an NSF Graduate Research Fellowship. E. Ratner is supported by a NASA Space Technology Research Fellowship.}
}

\begin{document}

\maketitle
\thispagestyle{empty}
\pagestyle{empty}

\begin{abstract}
\revision{Many problems in robotics involve multiple decision making agents. 
To operate efficiently in such settings, a robot must reason about the impact of its decisions on the behavior of other agents. 
Differential games offer an expressive theoretical framework for formulating these types of multi-agent problems. 
Unfortunately, most numerical solution techniques scale poorly with state dimension and are rarely used in real-time applications. 
For this reason, it is common to predict the future decisions of other agents and solve the resulting decoupled, i.e., single-agent, optimal control problem.
This decoupling neglects the underlying interactive nature of the problem; however, efficient solution techniques do exist for broad classes of optimal control problems.
We take inspiration from one such technique, the iterative linear-quadratic regulator (ILQR), which solves repeated approximations with linear dynamics and quadratic costs.
Similarly, our proposed algorithm solves repeated linear-quadratic \emph{games}.
We experimentally benchmark our algorithm in several examples with a variety of initial conditions and show that the resulting strategies exhibit complex interactive behavior.
Our results indicate that our algorithm converges reliably and runs in real-time.
In a three-player, $14$-state simulated intersection problem, our algorithm initially converges in ${{<0.25}\,\textbf{s}}$. Receding horizon invocations converge in ${<50}$\,ms in a hardware collision-avoidance test.
}
\end{abstract}

\input{intro.tex}

\input{related_work.tex}

\input{problem.tex}

\input{framework.tex}

\input{examples.tex}

\input{conclusion.tex}

\section*{Acknowledgments}
The authors would like to thank Andrew Packard for his helpful insights on LQ games, as well as Forrest Laine, Chih-Yuan Chiu, Somil Bansal, Jaime Fisac, Tyler Westenbroek, and Eric Mazumdar for helpful discussions.

\printbibliography

\end{document}

%% file: notation.tex
\newcommand{\dyn}{f}
\newcommand{\dx}{\dot{x}}

\newcommand{\numplayers}{N}
\newcommand{\cost}{J}
\newcommand{\runningcost}{g}
\newcommand{\horizon}{T}
\newcommand{\playerset}{[\numplayers]}
\newcommand{\feedback}{\gamma}
\newcommand{\feedbackset}{\Gamma}
\newcommand{\stepsize}{\eta}
\newcommand{\thresh}{d}
\newcommand{\lane}{\ell}

\newcommand{\traj}{\xi}
\newcommand{\xdim}{n}
\newcommand{\udim}{m}
\newcommand{\tset}{[0, \horizon]}
\newcommand{\xset}{\mathbb{R}^\xdim}
\newcommand{\uset}[1]{\mathbb{R}^{\udim_{#1}}}

\newcommand{\bigoh}{\mathcal{O}}

\def\atan2{\operatorname{atan2}}

\newtheorem{lemma}{Lemma}

%% file: intro.tex
\section{Introduction}
\label{sec:intro}

\begin{figure}[t!]
  \centering
  \includegraphics[width=0.5\columnwidth]{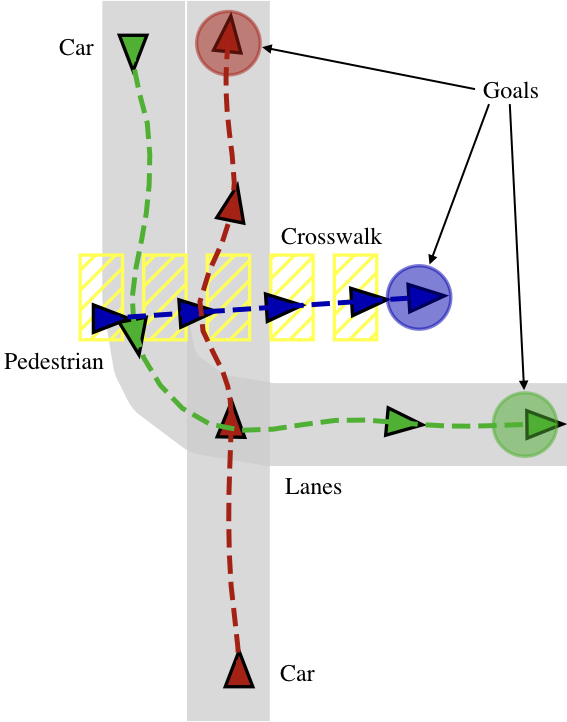}
  \caption{Demonstration of the proposed algorithm for a three-player general-sum game modeling an intersection. Two cars (red and green triangles) navigate the intersection while a pedestrian (blue triangle) traverses a crosswalk. Observe how both cars swerve slightly to avoid one another and provide extra clearance to the pedestrian.} 
  \label{fig:front_fig}
  \vspace{-2em}
\end{figure}

\revision{Many problems in robotics require an understanding of how multiple intelligent agents interact.
For example, in the intersection depicted in Fig.~\ref{fig:front_fig}, two cars and a pedestrian wish to reach their respective goals without colliding or leaving their lanes.
Successfully navigating the intersection requires either explicit, or perhaps implicit, coordination amongst the agents.
Often, these interactions are \emph{decoupled}, with each autonomous agent predicting the behavior of others and then planning an appropriate response.
This decoupling necessitates strong predictive assumptions on how agents' decisions impact one another.
Differential game theory provides a principled formalism for expressing these types of multi-agent decision making problems without requiring \emph{a priori} predictive assumptions.
}

Unfortunately, most classes of differential games have no analytic solution, and many numerical techniques suffer from the so-called ``curse of dimensionality'' \cite{bellman1956dynamic}.
Numerical dynamic programming solutions for general nonlinear systems have been studied extensively, though primarily in cases with \emph{a priori} known objectives and constraints which permit offline computation, such as automated aerial refueling \cite{ding2008reachability}.
Approaches such as \cite{herbert2017fastrack, fisac2018hierarchical} which separate offline game analysis from online operation are promising. Still, scenarios with more than two players remain extremely challenging, and the practical restriction of solving games offline prevents them from being widely used in many applications of interest, such as autonomous driving.

\revision{To simplify matters, decision making problems for multiple agents are often decoupled (see, e.g., \cite{ziebart2009planning, bai2015intention, schmerling2018multimodal}).
For example, the red car in Fig.~\ref{fig:front_fig} may wish to simplify its decision problem by \emph{predicting} the future motion of the other agents and \emph{plan reactively}. 
This simplification reduces the differential game to an optimal control problem, for which there often exist efficient solution techniques.
However, the decisions of the other agents \emph{will} depend upon what the red car chooses to do.
By ignoring this dependence, the red car is incapable of discovering strategies which exploit the reactions of others, and moreover, trusting in a nominal prediction---e.g., that the pedestrian will get of the way---may lead to unsafe behavior.
A differential game formulation of this problem, by contrast, explicitly accounts for the mutual dependence of all agents' decisions.}

\revision{We propose a novel \emph{local} algorithm for recovering interactive strategies in a broad class of differential games.
These strategies qualitatively resemble local Nash equilibria, though there are subtle differences.
By solving the underlying game we account for the fundamental interactive nature of the problem, and by seeking a local solution we avoid the curse of dimensionality which arises when searching for global Nash equilibria.
Our algorithm builds upon the iterative linear-quadratic regulator (ILQR) \cite{li2004iterative}, a local method used in smooth optimal control problems \cite{koenemann2015whole, kitaev2015physics, chen2017constrained}.
ILQR repeatedly refines an initial control strategy by efficiently solving approximations with linear dynamics and quadratic costs.
Like linear-quadratic (LQ) optimal control problems, LQ games also afford an efficient closed form solution \cite{basar1999dynamic}.
Our algorithm exploits this analytic solution to solve successive LQ approximations, and thereby finds a local solution to the original game in real-time.
For example, our algorithm initially solves the three-player 14-state intersection scenario of Fig.~\ref{fig:front_fig} in ${<0.25}$\,s, and receding horizon problems converge in ${<50}$\,ms in a hardware collision-avoidance test.}

%% file: related_work.tex
\section{Background \& Related Work}
\label{sec:related_work}


\subsection{General-sum games}
\label{subsec:general_sum_games}

Initially formulated in \cite{starr1969nonzero, starr1969further}, general-sum differential games generalize zero-sum games to model situations in which players have competing---but not necessarily opposite---objectives. Like zero-sum games, general-sum games are characterized by Hamilton-Jacobi equations \cite{starr1969nonzero} in which all players'  Hamiltonians are coupled with one other. 
Both zero-sum and general-sum games, and especially games with many players, are generally difficult to solve numerically.
However, efficient methods do exist for solving games with linear dynamics and quadratic costs, e.g. \cite{li1995lyapunov, basar1999dynamic}. Dockner et al. \cite{dockner1985tractable} also characterize classes of games which admit tractable \emph{open loop}, rather than \emph{feedback}, solutions.

\subsection{Approximation techniques}
\label{subsec:approximation_techniques}

While general-sum games may be analyzed by solving coupled Hamilton-Jacobi equations \cite{starr1969nonzero}, doing so requires both exponential time and computational memory.
A number of more tractable approximate solution techniques have been proposed for zero-sum games, many of which require linear system dynamics, e.g. \cite{kurzhanski00, kurzhanski02, greenstreet1999reachability, maidens13}, or decomposable dynamics \cite{chen2018decomposition}.
Approximate dynamic programming techniques such as \cite{bertsekas1996neuro} are not restricted to linear dynamics or zero-sum settings. Still, scalability to online, real-time operation remains a challenge.

Iterative best response algorithms form another class of approximate methods for solving general-sum games. Here, in each iteration every player solves (or approximately solves) the optimal control problem that results from holding other players' strategies fixed. 
This reduction to a sequence of optimal control problems is attractive; however, it can also be computationally inefficient.
Still recent work demonstrates the effectiveness of iterative best response in lane changes \cite{fisac2018hierarchical} and multi-vehicle racing \cite{wang2019game}.

Another similarly-motivated class of approximations involves changing the information structure of the game. For example, Chen et al. \cite{chen2015safe} solve a multi-player reach-avoid game by pre-specifying an ordering amongst the players and allowing earlier players to communicate their intended strategies to later players. Zhou et al. \cite{zhou2012general} and Liu et al. \cite{liu2014evasion} operate in a similar setting, but solve for open-loop conservative strategies.

\subsection{Iterative linear-quadratic (LQ) methods}
\label{subsec:ILQR}

Iterative LQ approximation methods are increasingly common in the robotics and control communities~\cite{chen2017constrained, koenemann2015whole, kitaev2015physics, van2014iterated, chen2017constrained}.
Our work builds directly upon the iterative linear-quadratic regulator (ILQR) algorithm \cite{li2004iterative, todorov2005generalized}. 


At each iteration, ILQR simulates the full nonlinear system trajectory, computes a discrete-time linear dynamics approximation and quadratic cost approximation, and solves a LQR subproblem to generate the next control strategy iterate. 
While structurally similar to ILQR, our approach solves a LQ game at each iteration instead of a LQR problem. This core idea is related to the sequential linear-quadratic method of \cite{mukai2000sequential, tanikawa2012local}, which is restricted to the two-player zero-sum context. In this paper, we show that LQ approximations can be applied in $N$-player, general-sum games. In addition, we experimentally characterize the quality of solutions in several case studies and demonstrate real-time operation.


%% file: problem.tex
\section{Problem Formulation}
\label{sec:problem}

We consider a $\numplayers$-player finite horizon general-sum differential game characterized by nonlinear system dynamics
\begin{align}
    \label{eqn:dynamics}
    \dx = \dyn(t, x, u_{1:\numplayers})\,,
\end{align}
where $x \in \xset$ is the state of the system, and $u_i \in \uset{i}, i \in \playerset \equiv \{1, \dots, \numplayers\}$ is the control input of player $i$, and $u_{1:\numplayers} \equiv (u_1, u_2, \dots, u_\numplayers)$. Each player has a cost function $\cost_i$ defined as an integral of running costs $\runningcost_i$. $\cost_i$ is understood to depend implicitly upon the state trajectory $x(\cdot)$, which itself depends upon initial state $x(0)$ and control signals $u_{1:\numplayers}(\cdot)$: 
\begin{align}
    \label{eqn:cost}
    \cost_i\big(&u_{1:\numplayers}(\cdot)\big) \triangleq \int_0^\horizon \runningcost_i\big(t, x(t), u_{1:\numplayers}(t)\big) dt, \forall i \in \playerset\,.
\end{align}

We shall presume that $\dyn$ is continuous in $t$ and continuously differentiable in $\{x, u_i\}$ uniformly in $t$. We shall also require $g_i$ to be twice differentiable in $\{x, u_i\}, \forall t$. 

Ideally, we would like to find time-varying state feedback control strategies $\feedback_i^{*} \in \feedbackset_i$ for each player $i$ which constitute a global Nash equilibrium for the game defined by \eqref{eqn:dynamics} and \eqref{eqn:cost}. Here, the strategy space $\feedbackset_i$ for player $i$ is the set of measurable functions $\feedback_i : \tset \times \xset \rightarrow \uset{i}$ mapping time and state to player $i$'s control input. Note that, in this formulation, player $i$ only observes the state of the system at each time and is unaware of other players' control strategies. With a slight abuse of notation $\cost_i(\feedback_1; \dots; \feedback_\numplayers) \equiv \cost_i\big(\feedback_1(\cdot, x(\cdot)), \dots, \feedback_\numplayers(\cdot, x(\cdot))\big)$, the global Nash equilibrium is defined as the set of strategies $\{\feedback_i\}$ for which the following inequalities hold (see, e.g., \cite[Chapter 6]{basar1999dynamic}):
\revision{\begin{equation}
\label{eqn:nash}
\begin{aligned}
 \cost_i^* &\triangleq \cost_i(\feedback_1^*; \dots; \feedback_{i-1}^*; \feedback_{i}^*; \feedback_{i+1}^*; \dots \feedback_\numplayers^*) \\
 &\le \cost_i(\feedback_1^*; \dots; \feedback_{i-1}^*; \feedback_{i}; \feedback_{i+1}^*; \dots \feedback_\numplayers^*), \forall i\in\playerset\,.
\end{aligned}
\end{equation}}
In \eqref{eqn:nash}, the inequalities must hold for all $\feedback_i \in \feedbackset_i, \forall i \in \playerset$.
Informally, a set of feedback strategies $(\feedback_1^*, \dots, \feedback_\numplayers^*)$ is a global Nash equilibrium if no player has a unilateral incentive to deviate from their current strategy.

Since finding a global Nash equilibrium is generally computationally intractable, recent work in adversarial learning \cite{mazumdar2019finding} and motion planning \cite{wang2019game, wangmingyu2019game} consider local Nash equilibria instead. 
Further, \cite{wang2019game, wangmingyu2019game} simplify the information structure of the game and consider open loop, rather than feedback, strategies.
Local Nash equilibria are characterized similarly to \eqref{eqn:nash}, except that the inequalities may only hold in a local neighborhood within the strategy space \cite[Definition 1]{ratliff2016characterization}.
\revision{In this paper, we shall seek a related type of equilibrium, which we describe more precisely in Section~\ref{subsec:convergence}.
Intuitively, we seek strategies which satisfy the \emph{global} Nash conditions \eqref{eqn:nash} for the limit of a sequence of \emph{local} approximations to the game. Our experimental results indicate that it does yield highly interactive strategies in a variety of differential games.}


%% file: framework.tex
\section{Iterative Linear-Quadratic Games}
\label{sec:framework}

We approach the $\numplayers$-player general-sum game with dynamics \eqref{eqn:dynamics} and costs \eqref{eqn:cost} from the perspective of classical LQ games. It is well known that Nash equilibrium strategies for finite-horizon LQ games satisfy coupled Riccati differential equations. 
These coupled Riccati equations may be derived by substituting linear dynamics and quadratic running costs into the generalized HJ equations \cite{starr1969further} and analyzing the first order necessary conditions of optimality for each player \cite[Chapter 6]{basar1999dynamic}.
These coupled differential equations may be solved approximately in discrete-time using dynamic programming \cite{basar1999dynamic}.
We will leverage the existence and computational efficiency of this discrete-time LQ solution to solve successive approximations to the original \emph{non}linear \emph{non}quadratic game.

\subsection{Iterative LQ game algorithm}
\label{subsec:ilq_game}
Our iterative LQ game approach proceeds in stages, as summarized in Algorithm~\ref{alg:ilq_game}. We begin with an initial state $x(0)$ and initial feedback control strategies $\{\feedback_i^0\}$ for each player $i$, and integrate the system forward (line~\ref{line:operating_pt} of Algorithm~\ref{alg:ilq_game}) to obtain the current trajectory iterate $\traj^k \equiv \{\hat x(t), \hat u_{1:\numplayers}(t)\}_{t \in [0, \horizon]}$.
Next (line~\ref{line:linearize}) we obtain a Jacobian linearization of the dynamics $\dyn$ about trajectory $\traj^k$. At each time $t \in [0, \horizon]$ and for arbitrary states $x(t)$ and controls $u_i(t)$ we define deviations from this trajectory $\delta x(t) = x(t) - \hat x(t)$ and $\delta u_i(t) = u_i(t) - \hat u_i(t)$. Thus equipped, we compute a continuous-time linear system approximation about $\traj^k$:
\begin{align}
\label{eqn:linearize}
\dot{\delta x}(t) \approx A(t) \delta x(t) + \sum_{i\in\playerset} B_i(t) \delta u_i(t),
\end{align}
where $A(t)$ is the Jacobian $D_{\hat x} \dyn\big(t, \hat x(t), \hat u_{1:\numplayers}(t)\big)$ and $B_i(t)$ is likewise $D_{\hat u_i} \dyn\big(t, \hat x(t), \hat u_{1:\numplayers}(t)\big)$.

\input{algorithm.tex}

We also obtain a quadratic approximation to the running cost $\runningcost_i$ for each player $i$ (see line~\ref{line:quadraticize} of Algorithm~\ref{alg:ilq_game})
\begin{align}
\label{eqn:quadraticize}
\runningcost_i\big(&t, x(t), u_{1:\numplayers}(t)\big) \approx \nonumber\\
&\runningcost_i\big(t, \hat x(t), \hat u_{1:\numplayers}(t)\big) + \frac{1}{2} \delta x(t)^T \left(Q_i(t) \delta x(t) + 2 l_i(t)\right) + \nonumber\\
&\frac{1}{2}\sum_{j\in\playerset} \delta u_j(t)^T \left(R_{ij}(t) \delta u_j(t) + 2 r_{ij}(t)\right),
\end{align}
where vector $l_i(t)$ is the gradient $D_{\hat x} \runningcost_i$, $r_{ij}$ is $D_{\hat u_j} \runningcost_i$, and matrices $Q_i$ and $R_{ij}$ are Hessians $D^2_{\hat x \hat x} \runningcost_i$ and $D^2_{\hat u_{j} \hat u_{j}} \runningcost_i$, respectively. \revision{We neglect mixed partials $D^2_{\hat u_j \hat u_k} g_i$ and $D^2_{\hat x \hat u_j} g_i$ as they rarely appear in cost structures of practical interest, although they could be incorporated if needed.} 

Thus, we have constructed a finite-horizon continuous-time LQ game, which may be solved via coupled Riccati differential equations \cite{basar1999dynamic, green2012linear}. This results in a new set of \emph{candidate} feedback strategies $\{\tilde \feedback_i^k\}$ which constitute a feedback (global) Nash equilibrium of the LQ game \cite{basar1999dynamic}. In fact, these feedback strategies are affine maps of the form:
\begin{align}
\label{eqn:affine_feedback}
\tilde \feedback_i^k\big(t, x(t)\big) = \hat u_i(t) - P_i^k(t) \delta x(t) - \alpha_i^k(t)\,,
\end{align}
with gains $P_i^k(t) \in \mathbb{R}^{\udim_i \times \xdim}$ and affine terms ${\alpha_i^k(t) \in \mathbb{R}^{\udim_i}}$.

However, we find that choosing $\feedback_i^k = \tilde \feedback_i^k$ often causes Algorithm~\ref{alg:ilq_game} to diverge because the trajectory resulting from $\{\tilde \feedback_i\}$ is far enough from the current trajectory iterate $\traj^k$ that the dynamics linearizations (Algorithm~\ref{alg:ilq_game}, line~\ref{line:linearize}) and cost quadraticizations (line~\ref{line:quadraticize}) no longer hold.
As in ILQR \cite{tassa2014control}, to improve convergence, we take only a small step in the ``direction'' of $\tilde \feedback_i^k$.\footnote{We also note that, in practice, it is often helpful to ``regularize'' the problem by adding scaled identity matrices $\epsilon I$ to $Q_i$ and/or $R_{ij}$.}
More precisely, for some choice of step size $\stepsize\in(0, 1]$, we set
\begin{align}
\label{eqn:affine_feedback_step}
\feedback_i^k\big(t, x(t)\big) = \hat u_i(t) - P_i^k(t) \delta x(t) - \stepsize \alpha_i^k(t)\,,
\end{align}
which corresponds to line~\ref{line:step} in Algorithm~\ref{alg:ilq_game}.
Note that at $t = 0$, $\delta x(0) = 0$ and $\feedback_i^k\big(0, x(0)\big) = \hat u_i(0) - \stepsize \alpha_i^k(0)$. Thus, taking $\stepsize = 0$, we have $\feedback_i^k\big(t, x(t)\big) = \hat u_i(t)$ (which may be verified recursively). That is, when $\eta = 0$ we recover the open-loop controls from the previous iterate, and hence $x(t) = \hat x(t)$.  Taking $\eta = 1$, we recover the LQ solution in \eqref{eqn:affine_feedback}. Similar logic implies the following lemma.
\begin{lemma}
\label{lemma:alpha_to_zero}
Suppose that trajectory $\xi^*$ is a fixed point of Algorithm~\ref{alg:ilq_game}, with $\eta \ne 0$. Then, the converged affine terms $\{\alpha_i^*(t)\}$ must all be identically zero for all time.
\end{lemma}

In ILQR, it is important to perform a line-search over step size $\stepsize$ to ensure a sufficient decrease in cost at every iteration, and thereby improve convergence (e.g., \cite{tassa2014control}). 
In the context of a noncooperative game, however, line-searching to decrease ``cost'' does not make sense, as costs $\{\cost_i\}$ may conflict.
For this reason, like other local methods in games (e.g., \cite{wang2019game}), our approach is not guaranteed to converge from arbitrary initializations.
In practice, however, we find that our algorithm typically converges for a fixed, small step size (e.g. $\stepsize = 0.01$). Heuristically decaying step size with each iteration $k$ or line-searching until $\|\traj^k - \traj^{k-1}\|$ is smaller than a threshold are also promising alternatives.
Further investigation of line-search methods in games is a rich topic of future research.

\emph{Note:}
Although we have presented our algorithm in continuous-time, in practice, we solve the coupled Riccati equations analytically in discrete-time via dynamic programming. Please refer to \cite[Corollary 6.1]{basar1999dynamic} for a full derivation. To discretize time at resolution $\Delta t$, we employ Runge-Kutta integration of nonlinear dynamics \eqref{eqn:dynamics} with a zero-order hold for control input over each time interval $\Delta t$. 

\subsection{Characterizing fixed points}
\label{subsec:convergence}

\revision{Suppose Algorithm~\ref{alg:ilq_game} converges to a fixed point $(\feedback_1^*, \dots, \feedback_\numplayers^*)$.
These strategies are the \emph{global} Nash equilibrium of a \emph{local} LQ approximation of the original game about the limiting operating point $\traj^*$.
While it is tempting to presume that such fixed points are also local Nash equilibria of the original game, this is not always true because converged strategies are only optimal for a LQ \emph{approximation} of the game at every time rather than the original game.
This approximation neglects higher order coupling effects between each player's running cost $g_i$ and other players' inputs $u_{j},j\ne i$.
These coupling effects arise in the game setting but \emph{not} in the optimal control setting, where ILQR converges to local minima.}

\subsection{Computational complexity and runtime}
\label{subsec:computational_complexity}

The per-iteration computational complexity of our approach is comparable to that of ILQR, and scales modestly with the number of players, $\numplayers$.
Specifically, at each iteration, we first linearize system dynamics about $\traj^k$. Presuming that the state dimension $\xdim$ is larger than the control dimension $\udim_i$ for each player, linearization requires computing $\bigoh(\xdim^2)$ partial derivatives at each time step (which also holds for ILQR). We also quadraticize costs, which requires $\bigoh(\numplayers \xdim^2)$ partial derivatives at each time step (compared to $\bigoh(\xdim^2)$ for ILQR). Finally, solving the coupled Riccati equations of the resulting LQ game at each time step has complexity $\bigoh(\numplayers^3 \xdim^3)$, which may be verified by inspecting \cite[Corollary 6.1]{basar1999dynamic} (for ILQR, this complexity is $\bigoh(\xdim^3)$).

Total algorithmic complexity depends upon the number of iterations, which we currently have no theory to bound. However, empirical results are extremely promising.
For the three-player 14-state game described in Section~\ref{subsec:intersection}, each iteration takes ${<8}$\,ms and the entire game can be solved from a zero initialization ($P_i^0(\cdot) = 0, \alpha_i^0(\cdot) = 0$) in ${<0.25}$\,s. Moreover, receding horizon invocations in a hardware collision-avoidance test can be solved in ${<50}$\,ms (Section~\ref{subsec:hardware}). All computation times are reported for single-threaded operation on a 2017 MacBook Pro with a 2.8 GHz Intel Core i7 CPU. For reference, the iterative best response scheme of \cite{wangmingyu2019game} reports solving a receding horizon two-player zero-sum racing game at $2$\,Hz, and the method of \cite{tanikawa2012local} reportedly takes several minutes to converge for a different two-player zero-sum example.
The dynamics and costs in both cases differ from those in Section~\ref{sec:examples} (or are not clearly reported); nonetheless, the runtime of our approach compares favorably.


%% file: algorithm.tex
\begin{algorithm}[t]
\KwIn{initial state $x(0)$, control strategies $\{\feedback_i^0\}_{i \in \playerset}$, time horizon $\horizon$, running costs $\{g_i\}_{i \in \playerset}$}
\KwOut{converged control strategies $\{\feedback_i^*\}_{i \in \playerset}$}

\For{iteration $k = 1, 2, \dots$}{
    $\traj^k \equiv \{\hat x(t), \hat u_{1:\numplayers}(t)\}_{t \in [0, t]} \leftarrow$\\ \Indp getTrajectory$\big(x(0), \{\feedback_i^{k-1}\}\big)$\nllabel{line:operating_pt}\;
    \Indm
    $\{A(t), B_i(t)\} \leftarrow$ linearizeDynamics$\big(\traj^k\big)$\nllabel{line:linearize}\;
    $\{l_i(t), Q_i(t), r_{ij}(t), R_{ij}(t)\} \leftarrow$ quadraticizeCost$\big(\traj^k\big)$\nllabel{line:quadraticize}\;
    $\{\tilde \feedback_i^k\} \leftarrow \textnormal{solveLQGame}\big($\\ \Indp$\{A(t), B_i(t), l_i(t), Q_i(t), r_{ij}(t), R_{ij}(t)\}\big)$\nllabel{line:solve_lq}\;
    \Indm
    $\{\feedback_i^k\} \leftarrow \textnormal{stepToward}\big(\{\feedback_i^{k-1}, \tilde \feedback_i^k\}\big)$\nllabel{line:step}\;
    \If{converged\nllabel{line:converged}}{
        \Return{$\{\feedback_i^k\}$}
    }
}
\caption{Iterative LQ Games \label{alg:ilq_game}}
\end{algorithm}

%% file: examples.tex
\section{Examples}
\label{sec:examples}

In this section, we demonstrate our algorithm experimentally in three-player noncooperative settings, both in software simulation and hardware.\footnote{Video summary available at \url{https://youtu.be/KPEPk-QrkQ8}.}

\subsection{Monte Carlo study}
\label{subsec:hallway}


\revision{We begin by presenting a Monte Carlo study of the convergence properties of Algorithm~\ref{alg:ilq_game}.
As we shall see, the solution to which Algorithm~\ref{alg:ilq_game} converges depends upon the initial strategy of each player, $\feedback_i^0$.
For clarity, we study this sensitivity in a game with simplified cost structure so that differences in solution are more easily attributable to coupling between players.}

\revision{Concretely, we consider a three-player ``hallway navigation'' game with time horizon $10$\,s and discretization $0.1$\,s.
Here, three people wish to interchange positions in a narrow hallway while maintaining at least $1$\,m clearance between one another.
We model each player $i$'s motion as:
\begin{equation}
\label{eqn:unicycle}
\begin{aligned}
    \dot p_{x, i} &= v_i \cos(\theta_i)\,,&\dot \theta_i &= \omega_i\,,\\
    \dot p_{y, i} &= v_i \sin(\theta_i)\,,&\dot v_i &= a_i\,,
\end{aligned}
\end{equation}
where $p_i := (p_{x, i}, p_{y, i})$ denotes player $i$'s position, $\theta_i$ heading angle, $v_i$ speed, and input $u_i := (\omega_i, a_i)$ yaw rate and longitudinal acceleration.
Concatenating all players' states into a global state vector $x := (p_{x, i}, p_{y, i}, \theta_i, v_i)_{i = 1}^3$, the game has 12 state dimensions and six input dimensions.}

\revision{We encode this problem with running costs $g_i$ \eqref{eqn:cost} expressed as weighted sums of the following:
\begin{align}
    \text{wall:}~&\mathbf{1}\{|p_{y, i}| > \thresh_{\text{hall}}\}(|p_{y, i}| - \thresh_{\text{hall}})^2 \label{eqn:wall_cost}\\    \text{proximity:}~&\mathbf{1}\{\|p_i - p_{j}\| < \thresh_{\text{prox}}\} (\thresh_{\text{prox}} - \|p_i - p_{j}\|)^2  \label{eqn:prox_cost}\\
    \text{goal:}~&\mathbf{1}\{t > T - t_{\text{goal}}\}\|p_i - p_{\text{goal}, i}\|^2  \label{eqn:goal_cost}\\
    \text{input:}~&u_i^T R_{ii} u_i
    \label{eqn:input_cost}
\end{align}
Here, $\mathbf{1}\{\cdot\}$ is the indicator function, i.e., it takes the value 1 if the given condition holds, and 0 otherwise. $\thresh_{\text{hall}}$ and $\thresh_{\text{prox}}$ denote threshold distances from hallway center and between players, which we set to $0.75$ and $1$\,m, respectively.
The goal cost is active only for the last $t_{\text{goal}}$ seconds, and the goal position is given by $p_{\text{goal}, i}$ for each player $i$.
Control inputs are penalized quadratically, with $R_{ii}$ a diagonal matrix.
The hallway is too narrow for all players to cross simultaneously without incurring a large proximity cost; hence, this proximity cost induces strong coupling between players' strategies.}

\begin{figure}
    \centering
   \includegraphics[width=0.8\linewidth]{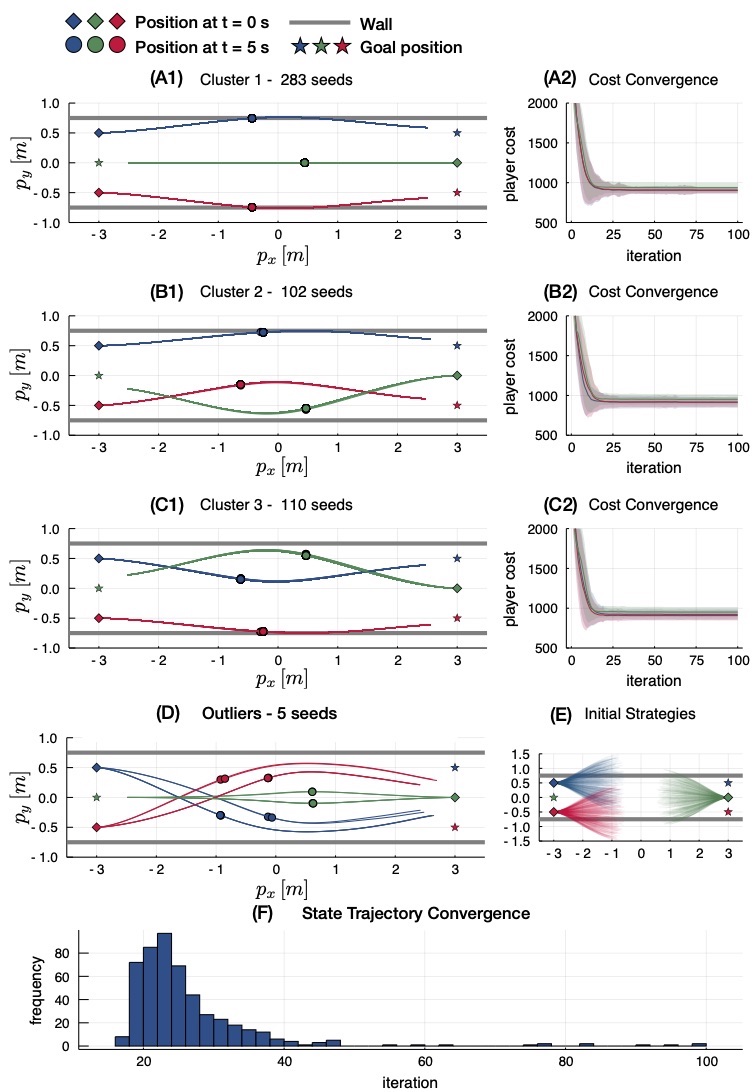}
    \caption{\revision{Monte Carlo results for a three-player hallway navigation game. (A1, B1, C1) Converged trajectories clustered by total Euclidean distance; each cluster corresponds to a qualitatively distinct mode of interaction. (A2, B2, C2) Costs for each player at each solver iteration. The shaded region corresponds to one standard deviation. (D) Several converged trajectories did not match a cluster (A-C). (E) Trajectories resulting from 500 random initial strategies. (F) Histogram of iterations until state trajectory has converged.}}
    \label{fig:hallway}
    \vspace{-2em}
\end{figure}

\revision{Fig.~\ref{fig:hallway} displays the results of our Monte Carlo study.
We seed Algorithm~\ref{alg:ilq_game} with 500 random sinusoidal open-loop initial strategies, which correspond to the trajectories shown in Fig.~\ref{fig:hallway}(E).
From each of these initializations, we run Algorithm~\ref{alg:ilq_game} for 100 iterations and cluster the resulting trajectories by Euclidean distance.
As shown in Fig.~\ref{fig:hallway}(A1, B1, C1), these clusters correspond to plausible modes of interaction; in each case, one or more players incur slightly higher cost to make room for the others to pass.
Beside each of these clusters in Fig.~\ref{fig:hallway}(A2, B2, C2), we also report the mean and standard deviation of each player's cost at each solver iteration.
As shown in Fig.~\ref{fig:hallway}(F), state trajectories converge within an $\ell_{\infty}$ tolerance of $0.01$ in well under 100 iterations.}

\revision{In these 500 random samples, only 6 did not converge and had to be resampled, and 5 converged to trajectories which were outliers from the clusters depicted in Fig.~\ref{fig:hallway}(A-C). These outliers are shown in Fig.~\ref{fig:hallway}(D).
We observe that, in these 5 cases, the players come within $0.5$\,m of one another.}


\subsection{Three-player intersection}
\label{subsec:intersection}

\begin{figure}
  \centering
  \includegraphics[width=0.9\linewidth,trim=0 12cm 0 0, clip=true]{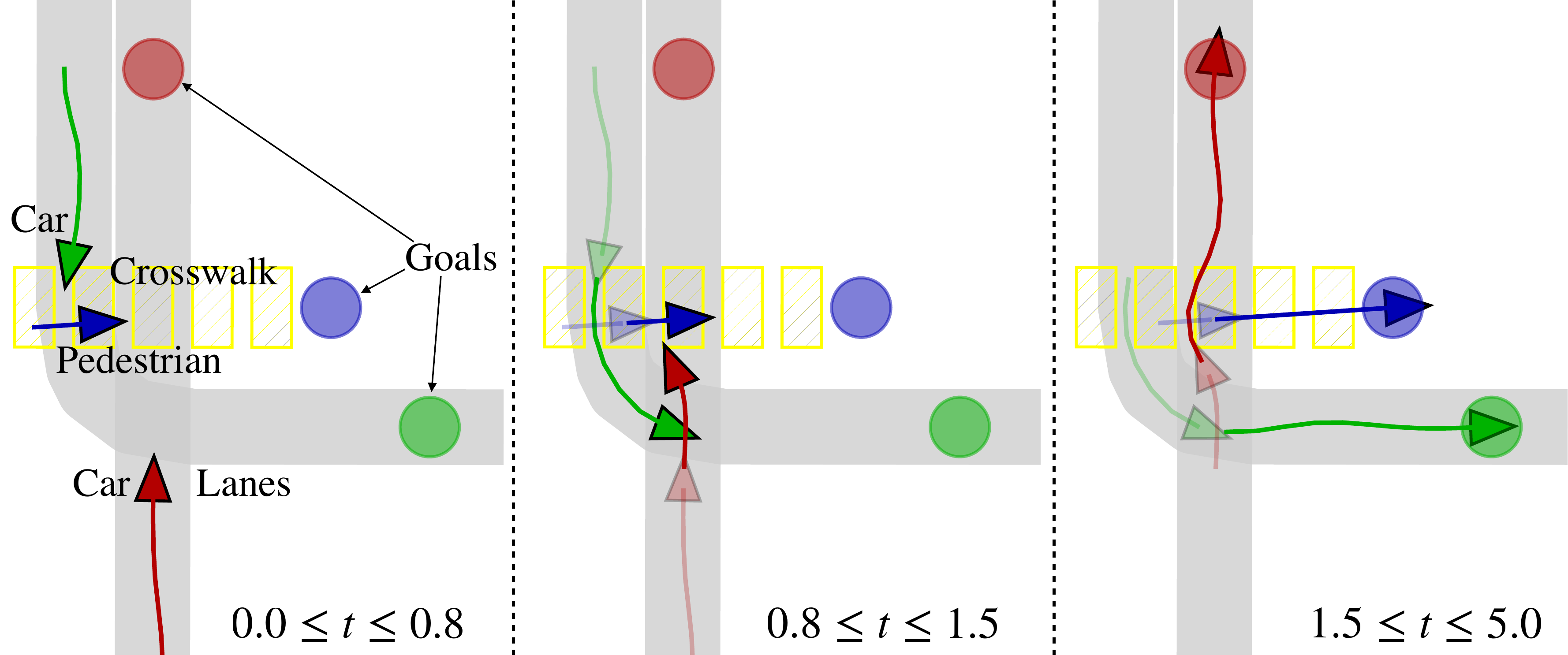}
  \caption{Three-player intersection game.
  (Left) Green car seeks the lane center and then swerves slightly to avoid the pedestrian. (Center) Red car accelerates in front of the green car and slows slightly to allow the pedestrian to pass. (Right) Red car swerves left to give pedestrian a wide berth.} 
  \label{fig:three_player_game_example}
  \vspace{-2em}
\end{figure}

Next, we consider a more complicated game intended to model traffic at an intersection.
As shown in Fig.~\ref{fig:three_player_game_example}, we consider an intersection with two cars and one pedestrian, all of which must cross paths to reach desired goal locations.
We use a time horizon of $5$\,s with discretization of $0.1$\,s, and Algorithm~\ref{alg:ilq_game} terminates in under $0.25$\,s.

\begin{figure*}
    \centering
    
    \includegraphics[trim={0 4.25cm 0cm 1.25cm}, clip, width=0.95\textwidth]{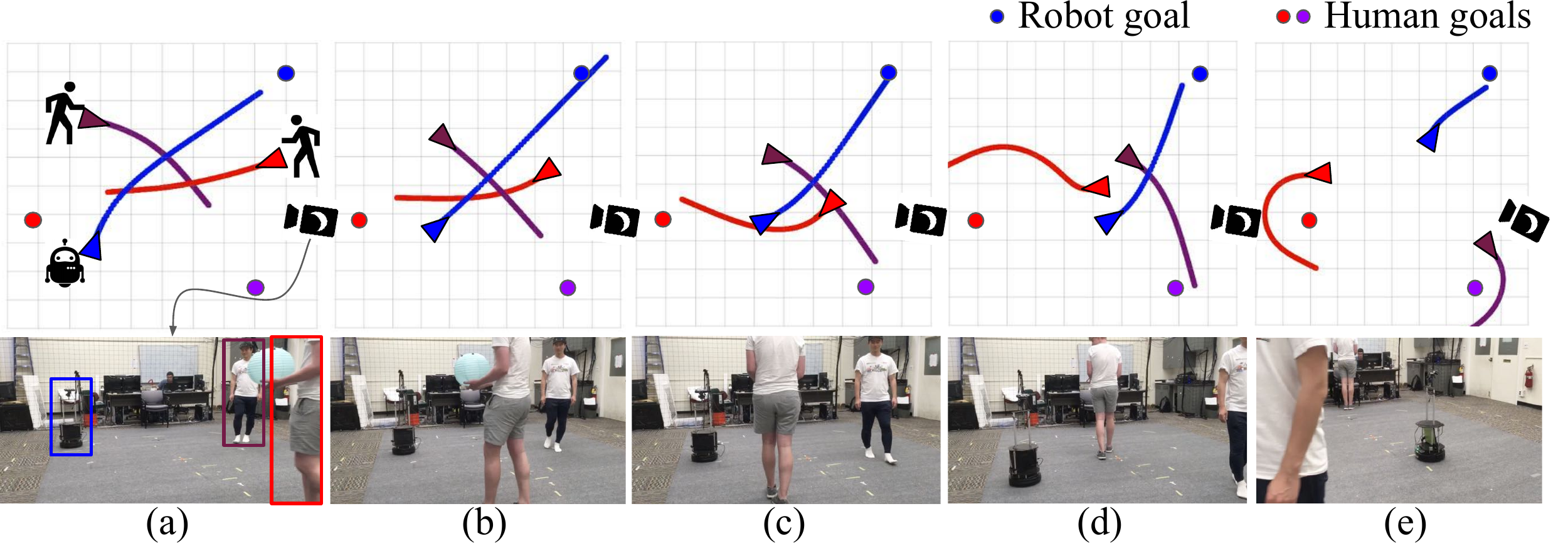}
    \caption{Time-lapse of a hardware demonstration of Algorithm~\ref{alg:ilq_game}. We model the interaction of a ground robot (blue triangle) and two humans (purple and red triangles) using a differential game in which each agent wishes to reach a goal location while maintaining sufficient distance from other agents. Our algorithm solves receding horizon instantiations of this game in real-time, and successfully plans and executes interactive collision-avoiding maneuvers. Planned (and predicted) trajectories are shown in blue (robot), purple, and red (humans).}
    \label{fig:hardware_time_lapse}
    \vspace{-2em}
\end{figure*}

We model the pedestrian's dynamics as in \eqref{eqn:unicycle} and each cars $i$'s dynamics as follows:
\begin{equation}
\label{eqn:bicycle}
\begin{aligned}
\dot p_{x, i} &= v_i \cos(\theta_i)\,, & \dot \theta_i &= v_i \tan(\phi_i) / L_i, & \dot \phi_i &= \psi_i\, & & \\
\dot p_{y, i} &= v_i \sin(\theta_i)\,, &\dot v_i &= a_i\,, & &
\end{aligned}
\end{equation}
where the state variables are as before \eqref{eqn:unicycle} except for front wheel angle $\phi_i$. $L_i$ is the inter-axle distance, and input $u_i := (\psi_i, a_i)$ is the front wheel angular rate and longitudinal acceleration, respectively.
Together, the state of this game is 14-dimensional.

\revision{The running cost for each player $i$ are specified as weighted sums of \eqref{eqn:prox_cost}--\eqref{eqn:input_cost}, and the following:
\begin{align}
    \text{lane center:}~&d_\lane(p_i)^2\\
    \text{lane boundary:}~&\mathbf{1}\{d_\lane(p_i) > \thresh_{\text{lane}}\}(\thresh_{\text{lane}} - d_\lane(p_i))^2\\
    \text{nominal speed:}~&(v_i - v_{\text{ref}, i})^2\\
    \text{speed bounds:}~&\mathbf{1}\{v_i > \overline{v}_i\}(v_i - \overline{v}_i)^2\nonumber \\&+\mathbf{1}\{v_i < \underline{v}_i\}(\underline{v}_i - v_i)^2
\end{align}
Here, $\thresh_{\text{lane}}$ denotes the lane half-width,  and $d_\lane(p_i) := \min_{p_\lane \in \lane} \|p_\lane - p_i\|$ measures player $i$'s distance to lane centerline $\lane$. 
Speed $v_i$ is penalized quadratically away from a fixed reference $v_{\text{ref}, i}$ also outside limits $\underline{v}_i$ and $\overline{v}_i$.}


Fig.~\ref{fig:three_player_game_example} shows a time-lapse of the converged solution identified by Algorithm~\ref{alg:ilq_game}.
These strategies exhibit non-trivial coordination among the players as they compete to reach their goals efficiently while sharing responsibility for collision-avoidance.
\revision{Such competitive behavior would be difficult for any single agent to recover from a decoupled, optimal control formulation.}
Observe how, between $0 \le t \le 0.8$\,s (left), the green car initially seeks the lane center to minimize its cost, but then turns slightly to avoid the pedestrian (blue).  Between $0.8 \le t \le 1.5$\,s (center), the red car turns right to pass in front of the green car, and then slows and begins to turn left to give the pedestrian time to cross. Finally (right), the red car turns left to give the pedestrian a wide berth.

\subsection{Receding horizon motion planning}
\label{subsec:hardware}

\revision{Differential games are appropriate in a variety of applications including multi-agent modeling and coordinated planning.
Here we present a proof-of-concept for their use in single-agent planning in a dynamic environment. 
In this setting, a single robot operates amongst multiple other agents whose true objectives are unknown.
The robot models these objectives and formulates the interaction as a differential game.
Then, crucially, the robot re-solves the differential game along a receding time horizon to account for possible deviations between the other agents' decisions and those which result from the game solution.}

We implement Algorithm~\ref{alg:ilq_game} in C++\footnote{Code available at: \url{github.com/HJReachability/ilqgames}} within the Robot Operating System (ROS) framework, and evaluate it in a real-time hardware test
onboard a TurtleBot 2 ground robot, in a motion capture room with two human participants.
The TurtleBot wishes to cross the room while maintaining ${>1\,\text{m}}$ clearance to the humans, and it models the humans likewise. 
We model the TurtleBot dynamics as \eqref{eqn:unicycle} and humans likewise but with constant speed $v_i$, i.e.:
\begin{equation}
\label{eqn:dubins}
\begin{aligned}
\dot p_{x, i} &= v_i \cos(\theta_i)\,, &\dot p_{y, i} &= v_i \sin(\theta_i)\,, &\dot \theta_i &= \omega_i\,.
\end{aligned}
\end{equation}

We use a similar cost structure as in Section~\ref{subsec:intersection}, and initialize Algorithm~\ref{alg:ilq_game} with all agents' strategies identically zero (i.e., $P_i^0(\cdot), \alpha_i^0(\cdot) \equiv 0$).
We re-solve the game in a $10$\,s receding horizon with time discretization of $0.1$\,s, and warm-start each successive receding horizon invocation with the previous solution.
Replanning every $0.25$\,s, Algorithm~\ref{alg:ilq_game} reliably converges in under $50$\,ms. We gather state information for all agents using a motion capture system.
Fig.~\ref{fig:hardware_time_lapse} shows a time-lapse of a typical interaction. 

Initially, in frame (a) Algorithm~\ref{alg:ilq_game} identifies a set of strategies which steer each agent to their respective goals while maintaining a large separation.
Of course, the human participants do not actually follow these precise trajectories; hence later receding horizon invocations converge to slightly different strategies. 
In fact, between frames (c) and (d) the red participant makes an unanticipated sharp right-hand turn, which forces the (blue) robot to stay to the right of its previous plan and then turn left in order to maintain sufficient separation between itself and both humans.
We note that our assumed cost structure models all agents as wishing to avoid collision. Thus, the resulting strategies may be less conservative than those that would arise from a non-game-theoretic motion planning approach. 


%% file: conclusion.tex
\section{Discussion}
\label{sec:conclusion}

We have presented a novel algorithm for finding local solutions in multi-player general-sum differential games.
Our approach is closely related to the iterative linear-quadratic regulator (ILQR) \cite{li2004iterative}, and offers a straightforward way for optimal control practitioners to directly account for multi-agent interactions via differential games. 
\revision{We performed a Monte Carlo study which demonstrated the reliability of our algorithm and its ability to identify complex interactive strategies for multiple agents.
These solutions display the competitive behavior associated with local Nash equilibria, although there are subtle differences.}
We also showcased our method in a three-player 14-dimensional traffic example, and tested it in real-time operation in a hardware robot navigation scenario, following a receding time horizon.

There are several other approaches to identifying local solutions in differential games, such as iterative best response \cite{wang2019game}.
We have shown the computational efficiency of our approach. 
However, quantitatively comparing the solutions identified by different algorithms is challenging due to \revision{differences in equilibrium concept, information structure (feedback vs. open loop), and implementation details.}
Furthermore, in arbitrary general-sum games, different players may prefer different equilibria.
Studying the qualitative differences in these equilibria is an important direction of future research. 

Although our experiments show that our algorithm converges reliably, we have no \emph{a priori} theoretical guarantee of convergence from arbitrary initializations. 
Future work will seek a theoretical explanation of this empirical property. 




